\documentclass[aps,reprint,twocolumn,pre,showpacs,amsmath,superscriptaddress,longbibliography]{revtex4-2}
\usepackage[dvipdfmx]{graphicx}
\usepackage[usenames,dvipsnames]{xcolor}
\usepackage{amsfonts,amsbsy,color}
\usepackage{amsmath,amssymb}
\usepackage{mathtools}
\usepackage[hidelinks]{hyperref}
\hypersetup{
  colorlinks   = true, 
  urlcolor     = blue, 
  linkcolor    = blue, 
  citecolor   = blue 
}

\newcommand{\bR}{\mathbb{R}}

\newcommand{\bZ}{\mathbb{Z}}

\newcommand{\calL}{\mathcal{L}}

\newcommand{\calN}{\mathcal{N}}

\newcommand{\calP}{\mathcal{P}}
\newcommand{\calQ}{\mathcal{Q}}

\newcommand{\calS}{\mathcal{S}}

\newcommand{\eq}{\mathrm{eq}}
\newcommand{\BZ}{\mathrm{BZ}_1}
\newcommand{\LM}{\mathrm{LM}}

\newcommand{\rev}{\mathrm{rev}}
\newcommand{\irr}{\mathrm{irr}}
\newcommand{\Tr}{\mathrm{Tr}}

\begin{document}
\title{Microscopic derivation of nonlinear fluctuating hydrodynamics for crystalline solid}

\author{Ken Hiura}
\email{ken.hiura@ubi.s.u-tokyo.ac.jp}
\affiliation {
Universal Biology Institute, The University of Tokyo, Tokyo 113-0033, Japan}

\date{\today}

\begin{abstract}
    We present a microscopic derivation of the nonlinear fluctuating hydrodynamic equation for a homogeneous crystalline solid from the Hamiltonian description of a many-particle system. We propose a microscopic expression of the displacement field that correctly generates the nonlinear elastic properties of the solid and find the nonlinear mode-coupling terms in reversible currents that are consistent with the phenomenological equation. The derivation relies on the projection onto the coarse-grained fields including the displacement field, the long-wavelength expansion, and the stationarity condition of the Fokker--Planck equation. 
\end{abstract}

\maketitle

\section{Introduction}


Fluctuating hydrodynamics is a universal theory for the slow variables that describes the long-time and large-distance behavior of dynamical fluctuations in many-body systems. The nonlinear couplings between fluctuating slow modes lead to non-trivial predictions in macroscopic transport phenomena that are difficult to understand analytically only on the basis of microscopic descriptions, e.g., the divergence of the transport coefficients in low-dimensional systems, the long-time tail in time correlation functions \cite{AlderWainwright1970,PomeauResibois1975}, and dynamic critical phenomena \cite{Kawasaki1970,HohenbergHalperin1977}. In addition to equilibrium systems, fluctuating hydrodynamics provides a useful starting point to study fluctuations in nonequilibrium steady states \cite{RonisProcaccia1982,Spohn1983,LutskoDufty1985}.


There have been various attempts for a long time to derive the hydrodynamic equation from a microscopic mechanical model \cite{IrvingKirkwood1950,Green1954,Mori1958,McLennan1960,ZubarevMorozovRople1997,ZubarevMorozov1983,Sasa2014,Hongo2019,SaitoHongoDharSasa2021}. When one derives the hydrodynamic equation, the slowly varying degrees of freedom in the system need to be identified. For three-dimensional simple fluids, the five locally conserved density fields, i.e., the number density, the three components of the momentum density, and the energy density, constitute a complete set of slow variables \cite{ZubarevMorozovRople1997,Sasa2014}. In crystalline solids at low temperatures, however, the translation invariance is spontaneously broken and the associated Nambu--Goldstone mode, the displacement field, must be included in relevant slow variables \cite{MartinParodiPershan1972,FlemingCohen1976,ChaikinLubensky1995}. These additional modes play an essential role in understanding the elastic properties of solids \cite{ChaikinLubensky1995,LandauLifshitzKosevichPitaevskii1986}. The purpose of this paper is to derive the hydrodynamic equation for crystals with the displacement field from a microscopic model.


In this paper, we derive the nonlinear fluctuating hydrodynamic equation for homogeneous crystalline solids in equilibrium from the Hamiltonian description of a many-particle system in a clear and compact manner. The work in this paper bears a lot of similarities with the previous work \cite{SzamelErnst1993,Szamel1997,MiserezGangulyHaussmannFuchs2022,MabillardGaspard2021,Haussmann2022}. The program for the microscopic derivation of the hydrodynamic equation for crystals was initiated by Szamel and his co-workers \cite{SzamelErnst1993,Szamel1997} and further developed recently by Refs. \cite{MiserezGangulyHaussmannFuchs2022,MabillardGaspard2021}. While these previous works used the microscopic definition of the displacement field proposed in \cite{SzamelErnst1993}, we show that the definition is not justified beyond the linear response regime. In this paper, to take the nonlinear elastic properties into account, we propose an alternative definition of the displacement field that correctly generates nonlinear hydrodynamics. By applying the projection operator method to the coarse-grained fields including the displacement field, we obtain the nonlinear fluctuating hydrodynamic equation for crystals. While Ref. \cite{Haussmann2022} applied a similar approach for deriving the nonlinear fluctuating hydrodynamics, the consistency with the phenomenological hydrodynamic equation is still unclear. We show that our resulting equation contains the nonlinear coupling terms and they are consistent with the phenomenological hydrodynamic equation. In parciular, the expression for the momentum current density (\ref{eq:momentumcurrent}), which is shown to be consistent with the phenomenological hydrodynamics, is not found in Ref. \cite{Haussmann2022}. Contrary to the previous works for linear hydrodynamics \cite{SzamelErnst1993,Szamel1997,MiserezGangulyHaussmannFuchs2022,MabillardGaspard2021}, our result presents a useful starting point for further analysis of nonlinear phenomena, e.g., the renormalization effect of the transport coefficients \cite{ForsterNelsonStephen1977}.

The paper is organized as follows. In Section \ref{sec:CG} we identify a relevant set of slow variables for crystalline solids and explain the coarse-graining procedure based on the projection operator method. Coarse-graining with the time-scale separation provides a formal expression for the nonlinear Langevin equation for hydrodynamic modes. In Section \ref{sec:current}, we determine the reversible and irreversible currents using the stationarity condition and the long-wavelength expansion. As a result, we obtain the explicit forms of reversible-mode coupling terms in terms of thermodynamic quantities and the Green--Kubo formula for the bare transport coefficients. 

\section{Coarse-graining procedure}
\label{sec:CG}

\subsection{Setup and conventions}
\label{subsec:setup}

For a set of lattice unit vectors $\{ \bar{f}^a : a = 1, 2, 3 \}$, we define the three-dimensional box $B_{L} = \{ \sum_{a=1}^{3} m_a \bar{f}^a : m_a \in [0, L) \}$ with side length $L$. We consider $N$ particles in $B_{L}$ and let $r_i$ and $p_i$ be the position and momentum of the $i$-th particle with mass $m$, respectively. We use $\Gamma = (r_i, p_i)_{i=1}^{N}$ to denote a microscopic configuration. A periodic boundary condition is imposed, and the particles interact via a translation-invariant potential $\Phi (r) = \sum_{n \in \bZ^3} \varphi ( r - L \sum_{a=1}^{3} n_a \bar{f}^a)$, where $\varphi$ is a short-ranged central potential function. The Hamiltonian of the system is $\hat{H} (\Gamma) \coloneqq \sum_{i=1}^{N} \hat{h}_i$ with
\begin{align}
 \hat{h}_i \coloneqq  \frac{p_{ia} p_{ia}}{2m} + \frac{1}{2} \sum_{j ( \neq i)} \Phi (r_i - r_j).
\end{align}
In this paper, the summation over repeated coordinate indices $a, b, c, \dots$ are implicitly assumed. We use $\Gamma_t$ to denote the solution at time $t$ of the Hamilton equation for an initial configuration $\Gamma$. The symbol $\hat{\cdot}$ on a variable indicates that it is a function on the phase space. We also simply use $f$ and $g$ to denote a vector $(f_a)_{a=1}^3$ and a tensor $(g_{ab})_{a,b=1}^3$, respectively. The Boltzmann constant is set to unity.

\subsection{Slow modes in crystalline solid}
\label{subsec:lcq}

In the crystalline phase, there are eight slow modes consisting of five locally conserved quantities and three components of the displacement field. First, we explain the five conserved quantities, which are the particle number, the three momentum components, and the energy. We define the corresponding empirical density fields, $\hat{n} (r, \Gamma) \coloneqq \sum_i \delta (r - \hat{r}_i)$, $\hat{\pi}_a (r, \Gamma) \coloneqq \sum_i \hat{p}_{ia} \delta (r - \hat{r}_i)$, and $\hat{e} (r, \Gamma) \coloneqq \sum_i \hat{h}_i \delta (r- \hat{r}_i)$, for $r \in \bR^3$. Since we impose the periodic boundary condition, the conserved density fields are periodic functions: $\hat{n} (r + L \sum_{a=1}^{3} m_a \bar{f}^a, \Gamma) = \hat{n} (r, \Gamma)$ for any $m \in \bZ^3$. Let $R_L \coloneqq \{ \sum_{a=1}^{3} m_a \bar{g}^a / L : m_a \in \bZ \}$ be the corresponding reciprocal space, where $\{ \bar{g}^a \}$ is a set of reciprocal unit vectors satisfying $\bar{f}^a_c \bar{g}_c^b = 2 \pi \delta_{ab}$. We introduce the Fourier components of the conserved density fields, $\{ \hat{n}_k, \hat{\pi}_{a,k}, \hat{e}_k : k \in R_L \}$. The Fourier transform convention in this paper is $\hat{n} (r, \Gamma) = \sum_{k \in R_L} e^{ik \cdot r} \hat{n}_k (\Gamma)$ with
\begin{align}
 \hat{n}_k (\Gamma) \coloneqq \frac{1}{V} \int_{B_L} e^{- ik \cdot r} \hat{n} (r, \Gamma) d^3r,
\end{align}
where $V$ is the volume of $B_L$. The conservation laws are expressed as continuity equations in terms of the locally conserved current densities $\{ \hat{j}^n_{a,k}, \hat{j}^{\pi}_{ab,k}, \hat{j}^{e}_{a,k} : k \in R_L \}$,
\begin{align}
 \partial_t \hat{n}_k (\Gamma_t) &= - ik_a \hat{j}^n_{a, k} (\Gamma_t),
 \\
 \partial_t \hat{\pi}_{a,k} (\Gamma_t) &= - ik_a \hat{j}^{\pi}_{ab, k} (\Gamma_t),
 \\
 \partial_t \hat{e}_{k} (\Gamma_t) &= - ik_a \hat{j}^{e}_{a, k} (\Gamma_t).
\end{align}
The microscopic expressions for locally conserved current densities are presented in Appendix \ref{subsec:microcurrent}.

We next consider the displacement field. A standard way to define the displacement field microscopically is to assign equilibrium lattice points to particles and then use the deviations of particle positions from the lattice points. This definition, however, misses the vacancy diffusion mode, which is present in crystals at finite temperatures \cite{MartinParodiPershan1972,FlemingCohen1976}. In this paper, we instead define the displacement field microscopically using the number density field without relying on the equilibrium lattice points, following Ref. \cite{SzamelErnst1993}. Let $D = \{ \sum_{a=1}^3 m_a l_a \bar{f}^a : m \in \bZ^3 \}$ and $G = \{ \sum_{a=1}^3 n_a \bar{g}^a / l_a : n \in \bZ^3 \}$ be a direct and reciprocal lattices with the lattice spacing $\{ l_a \}_{a=1}^3$, respectively. We suppose that the translational invariance is spontaneously broken at a sufficiently low temperature and the system is in a crystalline phase corresponding to the lattice $D$. The order parameter for the crystalline order is the set of Fourier components of the number density field in $G \backslash \{ 0 \}$, $\{ \hat{n}_g : g \in G \backslash \{ 0 \} \}$. To obtain a microscopic expression for the displacement field, we use a property in continuum mechanics that the displacement field must satisfy. For a macroscopic vector field $u$ characterizing the displacement from the equilibrium state, the number density field $n$ of the deformed crystal is
\begin{align} \label{eq:nu}
    n(r) = n_{\eq}(r - u(r)) \det [\delta_{ab} - \partial_a u_b],
\end{align}
where $n_{\eq} (r) = \sum_{g \in G} n_{\eq,g} e^{i g \cdot r}$ is the equilibrium number density field. This relation follows simply from the conservation of the particle number under the transformation $r \mapsto r - u(r)$. We remark that $n_{\eq}$ is not invariant under translations and rotations. We require the relation (\ref{eq:nu}) to hold between the microscopic displacement field $\hat{u}$ and the number density fluctuation $\delta \hat{n} = \hat{n} - n_{\eq}$. Since we study the fluctuations around the equilibrium state in this paper, the displacement field is assumed to be small. By expanding this relation with respect to $u$, we obtain
\begin{align} \label{eq:defdis}
    \hat{u}_{a,k} &= \hat{u}_{a,k}^0 + \sum_{q \in \BZ} \hat{u}_{a,q}^0 i(k_b-q_b) \hat{u}_{b,k-q}^0 +O((\hat{u}^0)^3),
\end{align}
with the linear displacement field
\begin{align} \label{eq:u0}
    \hat{u}^0_{a,k} \coloneqq (\calN^{-1})_{ab} \sum_{g \in G} ig_b n_{\eq,g}^* \delta \hat{n}_{k+g}.
\end{align}
Here, $\delta \hat{n}_k$ and $\hat{u}_{a,k}$ are, respectively, the Fourier components of the number density deviation and the displacement field, $\BZ$ is the first Brillouin zone, and $\calN_{ab} = \sum_{g \in G} g_a g_b |n_{\eq,g}|^2$. The derivation of Eq.~\eqref{eq:defdis} is presented in Appendix \ref{sec:disp}. Thus, the microscopic displacement field can be expressed in terms of short-wavelength fluctuations in number density around the reciprocal lattice points without assigning equilibrium points to particles. The displacement field (\ref{eq:defdis}) actually exhibits long-range order, and its long-wavelength components should be included in slow variables \cite{SzamelErnst1993}. We remark that Refs. \cite{SzamelErnst1993,Szamel1997,MiserezGangulyHaussmannFuchs2022,MabillardGaspard2021} used only the first term in Eq. ~\eqref{eq:defdis} as the displacement field, but the second term is needed to retain the nonlinear elastic properties as we will see later. The time evolution equation for the displacement field is expressed as
\begin{align}
    \partial_t \hat{u}_{a,k} (\Gamma_t) = - \hat{j}_{a,k}^u (\Gamma_t),
\end{align}
where $\hat{j}^u_{a,k}$ characterizes the decay rate of the displacement field. The microscopic expression of $\hat{j}^u_{a,k}$ is presented in Appendix \ref{sec:disp}.

We finally define a set of slow variables. There is a microscopic length scale $l$, which is of the order of the range of interaction and the lattice constants $\{ l_a \}$, and the macroscopic length scale $L$, which characterizes the system size. Assuming that these two length scales are well separated, we can take an intermediate length scale $\Lambda^{-1}$ satisfying $l \ll \Lambda^{-1} \ll L$. We expect that the fluctuating hydrodynamic description holds at this intermediate scale and that the form of the hydrodynamic equation is insensitive to the paricular choice of $\Lambda$. With this in mind, we choose $\hat{\alpha} = ( \hat{n}_k, \hat{\pi}_{k}, \hat{e}_k, \hat{u}_{k} : k \in R_L^{\Lambda} )$ as a set of relevant slow variables, where $R_L^{\Lambda} \coloneqq R_L \cap \{ |k| < \Lambda \}$. We use both $i$ and $\hat{\alpha}_i$ to designate an element in $\hat{\alpha}$, e.g., $\hat{\pi}_{1,k}$. Hereafter, the summation in Eq.~\eqref{eq:defdis} is restricted to $R_L^{\Lambda}$ for consistency.

We make a remark on the choice of slow variables in crystals. Ref. \cite{MiserezGangulyHaussmannFuchs2022} used the Fourie compoentns $\{ \delta \hat{n}_{k+g} : g \in F, k \in R^{\Lambda}_{L} \}$ of the density fluctuations, instead of the displacement field, as the additional slow variables. However, as we show later, simply adding the displacement field is sufficient to obtain the hydrodynamic equation consistent with the phenomenological theory. Therefore, in this paper, we do not include all the Fourier components of the density fluctuations as slow variables.

\subsection{Projection operator method}
\label{subsec:projection}

Dynamical fluctuations of the set of slow variables $\hat{\alpha}$ in equilibrium are characterized by the time evolution equation for the probability density of $\hat{\alpha}$. The probability density of $\hat{\alpha}$ is obtained by marginalizing the probability density function $\hat{\rho}_t$ at time $t$ on the phase space. That is, the marginal probability density $p_t$ at time $t$ for the slow variables is defined as
\begin{align}
 p_t (\alpha) \coloneqq \Tr [ \delta (\alpha - \hat{\alpha}) \hat{\rho}_t ],
\end{align}
where $\Tr [\cdot] = \sum_{\hat{N}} (\hat{N}!)^{-1} \int d\Gamma (\cdot)$ with $\hat{N} = V \hat{n}_0$, and
\begin{align}
 \delta (\alpha - \hat{\alpha}) &= \prod_{k \in R_L^{\Lambda}} \delta (n_k - \hat{n}_k) \delta (e_k - \hat{e}_k)
 \notag \\
 &\quad \times \prod_{a=1}^{3} \delta (\pi_{a,k} - \hat{\pi}_{a,k}) \delta (u_{a,k} - \hat{u}_{a,k}).
\end{align}
The time evolution of $\hat{\rho}_t$ is described by the Liouville equation
\begin{align} \label{eq:liouville}
 \partial_t \hat{\rho}_t = - \calL \hat{\rho}_t = \{ \hat{H}, \hat{\rho}_t \},
\end{align}
where we have introduced the Liouville generator $\calL = \{ \cdot, \hat{H} \}$. Our purpose is to obtain the closed form of the time evolution equation for $p_t$ from the Liouville equation (\ref{eq:liouville}). Following Ref. \cite{Zwanzig1961}, we introduce the projection operator $\calP$ acting on phase space functions as
\begin{align} \label{eq:zwanzig}
 (\calP \hat{f})(\Gamma) \coloneqq \Tr \left[ \hat{f} \frac{\delta (\hat{\alpha}(\Gamma) - \hat{\alpha})}{\Omega (\hat{\alpha} (\Gamma))} \right],
\end{align}
where $\Omega (\alpha) \coloneqq \mathrm{Tr}[ \delta (\alpha - \hat{\alpha} (\Gamma))]$ is the phase space volume of the set of configurations satisfying $\hat{\alpha} = \alpha$. Correspondingly, we define $\calQ \coloneqq 1 - \calP$. The operator $\calP$ is the projection onto the space of functions that depends on microscopic configurations only through the slow variables. If $\hat{f}$ is expressed as $\hat{f} (\Gamma) = f (\hat{\alpha}(\Gamma))$ for some function $f$, $\calP \hat{f} = \hat{f}$ and $\calQ \hat{f} = 0$. For simplicity, the initial probability density $\hat{\rho}_0$ is assumed to depend on $\Gamma$ only through $\hat{\alpha}$, i.e., $\calQ \hat{\rho}_0 = 0$. This assumption eliminates the effect of the relaxation of the fast degrees of freedom at the initial stage. Using the standard procedure in the projection operator method and assuming the Markovian approximation, we obtain the Fokker--Planck equation describing the equilibrium dynamics for the slow variables,
\begin{align} \label{eq:fp}
    \partial_t p_t &= - \sum_{l \in \hat{\alpha}} \frac{\partial}{\partial \alpha_l} \left( F_l p_t \right) + \sum_{l,m \in \hat{\alpha}} \frac{\partial}{\partial \alpha_l} \frac{\partial}{\partial \alpha_m} (L_{lm}^{(\mathrm{s})} p_t)
\end{align}
with the drift term
\begin{align}
    F_l  = V_l + \sum_m L_{lm} \frac{\partial \calS}{\partial \alpha_m} + \sum_m \frac{\partial L_{lm}}{\partial \alpha_m}.
\end{align}
Here, we have defined the local equilibrium velocity $V_l (\alpha) \coloneqq \langle \calL \hat{\alpha}_l \rangle^{\LM}_{\alpha}$. The average is taken with respect to the local microcanonical density $\hat{\rho}^{\LM}_{\alpha} \coloneqq \delta (\alpha - \hat{\alpha}) / \Omega (\alpha)$, i.e., $\langle \cdot \rangle^{\LM}_{\alpha} = \Tr [ (\cdot) \hat{\rho}^{\LM}_{\alpha}]$. We also define $\calS (\alpha) \coloneqq \ln \Omega (\alpha)$, the bare Onsager matrix
\begin{align} \label{eq:onsager}
    L_{lm}(\alpha) \coloneqq \int_0^{\infty} \langle (e^{\calQ \calL t} \calQ \calL \hat{\alpha}_l) ( \calQ \calL \hat{\alpha}_m) \rangle^{\LM}_{\alpha} dt,
\end{align}
and its symmetric part $L_{lm}^{(\mathrm{s})}(\alpha) = (L_{lm}(\alpha) + L_{ml}(\alpha))/2$. Thus, the short-time diffusion of slow variables is characterized by a certain type of time correlation function between fast-varying components of velocities $\{ \calQ \calL \hat{\alpha}_l = \calL \hat{\alpha}_l - V_l (\hat{\alpha}) \}$. The reversibility of the underlying dynamics leads to the reciprocal relation $L_{lm}(\alpha) = \epsilon_l \epsilon_m L_{ml}(\epsilon \alpha)$, where $\epsilon_l = +1$ or $-1$ according to the time-reversal symmetry of $\alpha_l$ and $\epsilon \alpha = (\epsilon_l \alpha_l)_{l}$. The Langevin equation corresponding to the Fokker--Planck equation (\ref{eq:fp}) is
\begin{align} \label{eq:langevin}
    d_t \alpha_l (t) = F_l (\alpha (t)) + \nu_l (t)
\end{align}
with the white Gaussian noise
\begin{align} \label{eq:noise}
    \langle \nu_l(t) \rangle = 0, \ \langle \nu_l(t) \nu_m(s) \rangle = 2 L_{lm}^{(\mathrm{s})} \delta (t-s),
\end{align}
where the brackets $\langle \cdot \rangle$ denote the average over the noise.

We remark that the projection operator introduced here is different from that in \cite{SzamelErnst1993,Szamel1997,MiserezGangulyHaussmannFuchs2022}. Refs. \cite{SzamelErnst1993,Szamel1997,MiserezGangulyHaussmannFuchs2022} used the projection onto the linear functions of slow variables. With this choice of the projection, we can obtain only the hydrodynamic equation where the drift term $F_l$ in Eq.~\eqref{eq:langevin} is linear with respect to $\alpha$. On the other hand, the projection operator (\ref{eq:zwanzig}) takes the nonlinear dependence of the phase space functions on the slow variables into consideration, and thus it allows us to obtain the nonlinear function $F_l$.

\section{Determination of currents}
\label{sec:current}

The drift terms and Onsager matrix in Eq.~\eqref{eq:fp} are obtained in principle from their microscopic expressions, but their connection with the thermodynamic quantities is not obvious. In this section, we determine the drift terms in terms of the thermodynamic quantities and the phenomenological transport coefficients.

\subsection{Local equilibrium drifts}

Although it is generally not easy to compute the local microcanonical averages, the local Galilean transformation and the conservation of entropy help us to find the thermodynamic expressions of the local microcanonical averages of the currents and decay rate in our case \cite{Sasa2014}. In this subsection, we often use the real-space expression of the slow variables, e.g., $\hat{n}(r) = \sum_{k \in R_L^{\Lambda}} e^{ik \cdot r} \hat{n}_k$, which should be distinguished from the fine-grained density field without the cutoff $\Lambda$.

For a slowly varying velocity field $v$ satisfying the periodic boundary condition, the local Galilean transformation maps a microscopic configuration $\Gamma =(r_i,p_i)_{i=1}^{N}$ into $\Gamma^v = (r_i^v, p_i^v)_{i=1}^{N}$ with $r_i^v = r_i$ and $p_i^v = p_i - mv (r_i)$. We introduce $\hat{f}^v(\Gamma) = \hat{f}(\Gamma^v)$ for a function $\hat{f}$ on the phase space. It is easy to see that in real space,
\begin{align} \label{eq:ncm}
    \hat{n} &= \hat{n}^v,
    \\ \label{eq:picm}
    \hat{\pi}_a &= \hat{\pi}_a^v + m \hat{n}^v v_a,
    \\ \label{eq:ecm}
    \hat{e} &= \hat{e}^v + \hat{\pi}_a v_a + \frac{m}{2} m \hat{n}^v |v|^2,
    \\ \label{eq:ucm}
    \hat{u}_a &= \hat{u}^{v}_a
\end{align}
and
\begin{align} \label{eq:jncm}
    \hat{j}^n_a &= \hat{j}^{n,v}_a + \hat{n}^v v_a,
    \\ \label{eq:jpicm}
    \hat{j}^{\pi}_{ab} &= \hat{j}^{\pi}_{ab} + \hat{\pi}_a^v v_b + \hat{\pi}_b^v v_a + m \hat{n}^v v_a v_b,
    \\ \label{eq:jecm}
    \hat{j}^e_a &= \hat{j}^{e,v}_a + \hat{j}^{\pi,v}_{ab} v_b + \hat{\pi}_a^v \frac{|v|^2}{2} 
    \notag \\
    &\quad + \left[ \hat{e}^v + \hat{\pi}_a^v v_a + \frac{m}{2} \hat{n}^v |v|^2 \right] v_a + O((\Lambda l)^2),
    \\ \label{eq:jucm}
    \hat{j}^{u}_a &= \hat{j}^{u,v}_a - v_a + (v \cdot \nabla) \hat{u}_a + O(\hat{u}^2 v).
\end{align}
For a set of values of slow variables $\alpha = (n_k, \pi_{k}, e_k, u_{k} : k \in R_L^{\Lambda})$, we define the coarse-grained velocity field $v$ through the relation $\pi_{a,k} = ( m n v_a)_k$ with the constraint $v_{a,k} = 0$ for $k \not\in R_L^{\Lambda}$. Hereafter, we use the velocity field defined this way and omit the dependence of $v$ on $\alpha$ from the notation. Eqs.~\eqref{eq:ncm}--\eqref{eq:ucm} imply $\hat{\rho}^{\LM}_{\alpha}(\Gamma) = \hat{\rho}^{\LM}_{\bar{\alpha}}(\Gamma^v)$ and $\langle \hat{f}^v \rangle^{\LM}_{\alpha} = \langle \hat{f} \rangle^{\LM}_{\bar{\alpha}}$ for $\bar{\alpha} = (n_k, 0, e_k - m(n|v|^2)_k/2, u_{k} : k \in R_L^{\Lambda})$. Using the latter relation with Eqs. (\ref{eq:jncm}) to (\ref{eq:jucm}) and the time-reversal symmetry, we obtain
\begin{align} \label{eq:lgjn}
    \langle \hat{j}^n_{a} \rangle^{\LM}_{\alpha} &= \frac{\pi_a}{m},
    \\ \label{eq:lgjpi}
    \langle \hat{j}^{\pi}_{ab} \rangle^{\LM}_{\alpha} &= p_{ab} + m n v_a v_b,
    \\ \label{eq:lgje}
    \langle \hat{j}^e_a \rangle^{\LM}_{\alpha} &= p_{ab} v_b + e v_a + O((\Lambda l)^2),
    \\ \label{eq:lgju}
    \langle \hat{j}^u_a \rangle^{\LM}_{\alpha} &= - v_a + (v \cdot \nabla) u_a + O(u^2 v)
\end{align}
with $p_{ab} \coloneqq \langle \hat{j}^{\pi}_{ab} \rangle^{\LM}_{\bar{\alpha}}$. 

To connect the local microcanonical average of the momentum current density $p_{ab}$ with the thermodynamic quantity, we use the equality
\begin{align} \label{eq:st}
    \sum_{l \in \hat{\alpha}} \left( \frac{\partial V_l}{\partial \alpha_l} + V_l \frac{\partial \calS}{\partial \alpha_l} \right) = 0,
\end{align}
which is equivalent to the stationarity condition for $\Omega$ of the Fokker--Planck equation (\ref{eq:fp}). The proof of Eq.~\eqref{eq:st} is in Appendx \ref{subsec:db}. Because $\calS (\alpha)$ is the logarithm of the phase space volume corresponding to the state specified by the long-wavelength components of the slow variables, we expect that $\calS$ can be approximated by the integral of the thermodynamic entropy density,
\begin{align} \label{eq:calS}
    \calS (\alpha) &= \calS_0 (\alpha) + O((\Lambda l)^2),
\end{align}
with
\begin{align} \label{eq:calS0}
    \calS_0 (\alpha) = \int_{B_L} s \left( e(r) - \frac{|\pi(r)|^2}{2m n(r)}, n(r), \varepsilon (r) \right) d^3r.
\end{align}
Here, $\varepsilon$ is the strain field defined by
\begin{align} \label{eq:defstrain}
    \varepsilon_{ab} = \frac{1}{2} \left( \partial_a u_b + \partial_b u_a - \partial_a u_c \partial_b u_c \right)
\end{align}
and $s(e^*, n, \varepsilon)$ is the thermodynamic entropy density as a function of the internal energy $e^* = e - |\pi|^2/2mn$, number density $n$, and strain $\varepsilon$. Because $\calS_0$ is of the order $O((L/l)^d)$ in the thermodynamic limit $L/l \to \infty$, the first term in Eq.~\eqref{eq:st} can be neglected in this limit. The resulting equality is exactly the same as Eq. (7) in Ref. \cite{Sasa2014} if the local microcanonical average is replaced by the local Gibbs average. Eq.~\eqref{eq:st} means that no entropy production dissipates if the slow variables evolve in time according to the drifts given by the local microcanonical averages of their velocities.

The thermodynamic relation in the crystalline phase is expressed as
\begin{align} \label{eq:thr}
    ds = \beta de - \beta \mu dn - \beta v_a d\pi_a - \phi_{ab} d\varepsilon_{ab},
\end{align}
where $\beta$ is the inverse temperature, $\mu$ is the chemical potential, and $\phi_{ab}$ is the conjugate variable of the strain $\varepsilon_{ab}$. In this paper, the pressure function $p$ is defined through the Euler relation $\beta p = s - \beta e + \beta \pi_a v_a + \beta \mu n + \beta \phi_{ab} \varepsilon_{ab}$. After substituting Eqs.~\eqref{eq:lgjn}--\eqref{eq:lgju} into the stationarity condition (\ref{eq:st}) with (\ref{eq:calS}) and some manipulation, we obtain
\begin{align} \label{eq:st2}
    &\int_{B_L} \left[ - \beta (r) (\delta p_{ab} (r)) (\partial_a v_b)(r)  + O((\delta \alpha)^4) \right] = 0
\end{align}
with
\begin{align} \label{eq:defpab}
    \delta p_{ab} &\coloneqq p_{ab} - p \delta_{ab} + \phi_{ab} + \phi_{cd} \varepsilon_{cd} \delta_{ab} 
    \notag \\
    &\qquad - \frac{1}{2} (\phi_{ac} \varepsilon_{bc} + \phi_{bc} \varepsilon_{ac}).
\end{align}
The term $O((\delta \alpha)^4)$ represents the fourth-order correction with respect to the deviations of the slow variables from the equilibrium values. The fields of thermodynamically conjugate variables in Eq.~\eqref{eq:st2} are given through the set of thermodynamic relations, e.g., $\beta (r) = \beta (e^*(r), n(r), \varepsilon (r))$. Because $\delta p_{ab}$ depends on the velocity field only though its norm $|v|$, Eq.~\eqref{eq:st2} implies $\delta p_{ab} = 0 + O((\delta \alpha)^3)$. Thus, the stationarity condition, which is equivalent to the conservation of entropy for the local equilibrium velocities, determines the thermodynamic expression of the local microcanonical average of the momentum current density up to the second-order fluctuations. 

We remark that the linear displacement field (\ref{eq:u0}) used in the previous studies \cite{SzamelErnst1993,Szamel1997,MiserezGangulyHaussmannFuchs2022,MabillardGaspard2021} does not produce the correct expression of the decay rate of the displacement field (\ref{eq:lgju}) beyond the linear response regime. Indeed, if we take Eq.~\eqref{eq:u0} as the displacement field, the local equilibrium average of the decay rate is
\begin{align}
    \langle \hat{j}^{u^0}_a \rangle^{\LM}_{\alpha} = - v_a + (\nabla \cdot v) u_a + (\nabla \cdot u) v_a + (v \cdot \nabla) u_a,
\end{align}
where $\hat{j}^{u^0}_a$ is the decay rate associated with the linear displacement field $\hat{u}^0$. The explicit definition of $\hat{j}^{u^0}_a$ is presented in Appendix \ref{sec:disp}. As we show in Appendix \ref{subsec:thermodynamics}, this expression for the decay rate is not consistent with the phenomenological result (\ref{eq:juph}) contrary to Eq.~\eqref{eq:lgju}. This inconsistency comes from the fact that the linear displacement field captures only the linear fluctuations of the number density, as suggested by Eq.~\eqref{eq:u0}.

\subsection{Gradient drift}

We next simplify the drift terms that are proportional to the thermodynamic forces. This simplification is achieved by expansion around the equilibrium state and long-wavelength expansion. First, because we consider dynamical fluctuations around equilibrium, the local microcanonical averages for the bare Onsager coefficients (\ref{eq:onsager}) are replaced by the equilibrium averages.,
\begin{align}
    L_{lm}(\alpha) \simeq \int_0^{\infty} \langle (e^{\calQ \calL t} \calQ \calL \hat{\alpha}_l)( \calQ \calL \hat{\alpha}_m) \rangle_{\eq} dt.
\end{align}
The brackets $\langle \cdot \rangle_{\eq}$ denote the average with respect to the grand-canonical distribution. Within this approximation, the bare Onsager coefficients are determined only by the environmental condition, and therefore the derivatives of the bare Onsager coefficients in Eq.~\eqref{eq:fp} vanish. Second, we take the long-wavelength limit of the bare Onsager cofficients. In this limit, the bare Onsager coefficient associated with the energy--energy coupling becomes
\begin{align}
    V L_{\hat{e}_k, \hat{e}_{-k}} \simeq k_a k_b L_{ab}^{ee}.
\end{align}
Here, we have defined the wavenumber-independent Onsager coefficient
\begin{align} \label{eq:Lee}
    L_{ab}^{ee} \coloneqq \lim_{V \to \infty} \frac{1}{V} \int_0^{\infty} \langle (e^{\calQ \calL t} \calQ \hat{J}^e_a)( \calQ \hat{J}^e_b) \rangle_{\eq}
\end{align}
with the spatial integral of the energy current density $\hat{J}^e_a = V \hat{j}^e_{a, 0}$. From Eq.~\eqref{eq:lgje}, the projected energy current $\calQ \hat{J}^e_a$ is given by
\begin{align}
    \calQ \hat{J}^e_a = \frac{1}{V} \int (\hat{j}^e_a - p_{ab}(\hat{e}^*, \hat{n}, \hat{\varepsilon}) \hat{v}_b - \hat{e} \hat{v}_a) d^3r.
\end{align}
Using $\partial \calS / \partial \alpha_m = \partial \calS_0 / \partial \alpha_m + O((\Lambda l)^2)$, we conclude that the gradient term is
\begin{align}
    L_{\hat{e}_k, \hat{e}_{-k}} \frac{\partial \calS}{\partial e_{-k}} \simeq k_a k_b L_{ab}^{ee} \beta_k,
\end{align}
where $\beta_k$ is the Fourier component of the inverse temperature field. The conventional choice of the transport coefficient in the energy--energy coupling is the heat conductivity $\kappa$ defined by
\begin{align}
    \kappa_{ab} \coloneqq \frac{L^{ee}_{ab}}{T_{\eq}^2},
\end{align}
where  $T_{\eq}$ is the equilibrium temperature. With this choice, the gradient term of the energy--energy coupling reproduces the conventional dissipative term, $- \partial_a (\kappa_{ab} \partial_b T)$, in real-space representation if we neglect $O((\delta \alpha)^2)$ terms. Thus, Eq.~\eqref{eq:Lee} gives the Green--Kubo formula connecting the bare transport coefficient with the time correlation function between the projected currents. Other Onsager coefficients are also approximated by their long-wavelength limits: $V L_{\hat{\pi}_{a,k}, \hat{\pi}_{b,-k}} \simeq k_c k_d L^{\pi_a \pi_b}_{cd}$, $V L_{\hat{e}_k, \hat{u}_{a,-k}} \simeq  ik_b L^{eu_a}_{b*}$, $V L_{\hat{u}_{a,k}, \hat{e}_{-k}} \simeq - ik_b L^{u_a e}_{*b}$, $V L_{\hat{u}_{a,k}, \hat{u}_{b,-k}} \simeq L^{u_a u_b}_{**}$, $V L_{\hat{e}_k, \hat{\pi}_{a,-k}} \simeq k_b k_c L^{e \pi_a}_{bc}$, $V L_{\hat{\pi}_{a,k}, \hat{e}_{-k}} \simeq k_b k_c L^{\pi_a e}_{bc}$, $V L_{\hat{u}_{a,k}, \hat{\pi}_{b,-k}} \simeq -ik_c L^{u_a \pi_b}_{*c}$, and $V L_{\hat{\pi}_{a,k}, \hat{u}_{b,-k}} \simeq ik_c L^{\pi_a u_b}_{c*}$, where
\begin{align}
    L^{\pi_a \pi_b}_{cd} &\coloneqq \lim_{V \to \infty} \frac{1}{V} \int_0^{\infty} \langle (e^{\calQ \calL t} \calQ \hat{J}^{\pi}_{ac})(\calQ \hat{J}^{\pi}_{bd}) \rangle_{\eq} dt,
    \\
    L^{eu_a}_{b*} &\coloneqq \lim_{V \to \infty} \frac{1}{V} \int_0^{\infty} \langle (e^{\calQ \calL t} \calQ \hat{J}^e_b)( \calQ \hat{J}^u_a) \rangle_{\eq} dt,
    \\
    L^{u_a e}_{*b} &\coloneqq \lim_{V \to \infty} \frac{1}{V} \int_0^{\infty} \langle (e^{\calQ \calL t} \calQ \hat{J}^u_a) (\calQ \hat{J}^e_b) \rangle_{\eq} dt,
    \\
    L^{u_a u_b}_{**} &\coloneqq \lim_{V \to \infty} \frac{1}{V} \int_0^{\infty} \langle (e^{\calQ \calL t} \calQ \hat{J}^u_a)(\calQ \hat{J}^u_b) \rangle_{\eq} dt,
    \\
    L^{e \pi_a}_{bc} &\coloneqq \lim_{V \to \infty} \frac{1}{V} \int_0^{\infty} \langle (e^{\calQ \calL t} \calQ \hat{J}^e_b) (\calQ \hat{J}^{\pi}_{ac}) \rangle_{\eq} dt,
    \\
    L^{\pi_a e}_{bc} &\coloneqq \lim_{V \to \infty} \frac{1}{V} \int_0^{\infty} \langle (e^{\calQ \calL t} \calQ \hat{J}^{\pi}_{ab}) ( \calQ \hat{J}^e_c) \rangle_{\eq} dt,
    \\
    L^{u_a \pi_b}_{*c} &\coloneqq \lim_{V \to \infty} \frac{1}{V} \int_0^{\infty} \langle (e^{\calQ \calL t} \calQ \hat{J}^u_a) ( \calQ \hat{J}^{\pi}_{bc}) \rangle_{\eq} dt,
    \\
    L^{\pi_a u_b}_{c*} &\coloneqq \lim_{V \to \infty} \frac{1}{V} \int_0^{\infty} \langle (e^{\calQ \calL t} \calQ \hat{J}^{\pi}_{ac}) (\calQ \hat{J}^u_b) \rangle_{\eq} dt,
\end{align}
with $\hat{J}^{\pi}_{ab} \coloneqq V \hat{j}^{\pi}_{ab,0}$ and $\hat{J}^u_a \coloneqq V \hat{j}^u_{a,0}$. The reciprocal relations are expressed as $L^{ee}_{ab} = L^{ee}_{ba}$, $L^{\pi_a \pi_b}_{cd} = L^{\pi_b \pi_a}_{dc}$, $L^{eu_a}_{b*} = L^{u_a e}_{*b}$, $L^{u_a u_b}_{**} = L^{u_b u_a}_{**}$, $L^{\pi_a e}_{bc} = - L^{e \pi_a}_{cb}$, and $L^{\pi_a u_b}_{c*} = - L^{u_a \pi_a}_{*a}$. Therefore, the independent transport coefficients are given by $\kappa_{ab}$, $\eta_{acbd} \coloneqq L^{\pi_a \pi_b}_{cd} / T_{\eq}$, $\xi_{ab} \coloneqq L^{eu_b}_{a*} / T_{\eq}$, $\zeta_{ab} \coloneqq L^{u_a u_b}_{**} / T_{\eq}$, $\chi_{abc} \coloneqq L^{e \pi_c}_{ab} / T_{\eq}$, and $\theta_{abc} \coloneqq L^{\pi_b u_c}_{a*} / T_{\eq}$ with $\kappa_{ab} = \kappa_{ba}$, $\eta_{abcd} = \eta_{cdab}$ and $\zeta_{ab} = \zeta_{ba}$. We remark that $L_{\hat{n}_k, \hat{\alpha}_l} = L_{\hat{\alpha}_l, \hat{n}_k} = 0$ for any $l \in \hat{\alpha}$ because $\calQ \calL \hat{n}_k = 0$. Consequently, the gradient terms $\sum_m L_{lm} \partial \calS / \partial \alpha_m$ for the energy $e_k$, the momentum $\pi_{a,k}$, and the displacement $u_{a,k}$ are given by, respectively,
\begin{align}
    &k_a k_b \kappa_{ab} T_{\eq}^2 \beta_k - k_a k_b \chi_{abc} v_{c,k} - k_a k_c \xi_{ab}  \phi_{bc,k},
    \\
    &k_b k_c \chi_{cba} T_{\eq} \beta_k -k_b k_c \eta_{abdc} v_{d,k}  - k_b k_d \theta_{bac}  \phi_{cd,k},
    \\
    & - ik_b \xi_{ba} T_{\eq} \beta_k - ik_c \theta_{cba} v_{b,k} + ik_c \zeta_{ab} \phi_{bc,k},
\end{align}
if $O((\delta \alpha)^2)$ terms are neglected. Here, $\phi_{ab,k}$ is the Fourier component of $\phi_{ab}$. Among these terms, the energy--momentum coupling and momentum--displacement coupling terms are reversible in that they do not contribute to the entropy production, because they have the same time-reversal symmetry as that of the currents in which they appear. The orders of these reversible gradient terms with respect to the wavenumbers are higher than those of the local equilibrium velocities in Eqs.~\eqref{eq:lgjn} to \eqref{eq:lgju}, which are all reversible. Therefore, these terms are often neglected \cite{Haussmann2022,MiserezGangulyHaussmannFuchs2022}. We follow that approximation in this paper. Furthermore, the associated Onsager coefficients have only the anti-symmetric parts, and therefore they do not appear in the noise amplitudes in Eq.~\eqref{eq:noise}.

The remaining terms are all irreversible and have only symmetric parts. Consequently, the covariances of the noises $\{ \nu_l \}$ in Eq.~\eqref{eq:noise} are given by
\begin{align}
    \langle \nu_{\hat{e}_k} \nu_{\hat{e}_{q}} \rangle &= 2 k_a k_b T_{\eq}^2 \kappa_{ab}  \frac{\delta_{k+q,0}}{V} \delta (t-s),
    \\
    \langle \nu_{\hat{e}_k} \nu_{\hat{\pi}_{a,q}}  \rangle &= 0,
    \\
    \langle \nu_{\hat{e}_k} \nu_{\hat{u}_{a,q}}  \rangle &= 2 ik_b T_{\eq} \xi_{ba} \frac{\delta_{k+q,0}}{V} \delta (t-s),
    \\
    \langle \nu_{\hat{\pi}_{a,k}} \nu_{\hat{\pi}_{b,q}} \rangle &= 2 k_c k_d T_{\eq} \eta_{acbd} \frac{\delta_{k+q,0}}{V} \delta (t-s),
    \\
    \langle \nu_{\hat{\pi}_{a,k}} \nu_{\hat{u}_{b,q}} \rangle &= 0,
    \\
    \langle \nu_{\hat{u}_{a,k}} \nu_{\hat{u}_{b,q}} \rangle &= 2 T_{\eq} \zeta_{ab} \frac{\delta_{k+q,0}}{V} \delta (t-s).
\end{align}

\subsection{Summary of hydrodynamic equation}

We summarize the nonlinear fluctuating hydrodynamics for crystals. The Langevin equation (\ref{eq:langevin}) becomes
\begin{align} \label{eq:n}
    &\partial_t n_k + ik_a \frac{\pi_{a,k}}{m} = 0,
    \\ \label{eq:pi}
    &\partial_t \pi_{a,k} + ik_b ( \sum_{q \in R_L^{\Lambda}} \pi_{a,q} v_{b,k-q} + p_{ab,k} 
    \notag \\
    &\qquad - \eta_{abdc} ik_c v_{d,k}  ) = \nu_{\hat{\pi}_{a,k}} ,
    \\ \label{eq:e}
    &\partial_t e_k + ik_a ( \sum_{a \in R_L^{\Lambda}} (e_q v_{a,k-q} + p_{ab,q} v_{b,k-q}) 
    \notag \\
    &\qquad + ik_b \kappa_{ab} T_{\eq}^2 \beta_k - \xi_{ab} ik_c \phi_{bc,k}) = \nu_{\hat{\pi}_{a,k}} ,
    \\ \label{eq:u}
    &\partial_t u_{a,k} + \sum_{q \in R_L^{\Lambda}} v_{b,k-q} iq_b u_{a,q} 
    \notag \\
    &\qquad = v_{a,k} - ik_b \xi_{ba} T_{\eq} \beta_k + \zeta_{ab} ik_c \phi_{bc,k} + \nu_{\hat{u}_{a,k}},
\end{align}
where $p_{ab,k}$ is the Fourier component of
\begin{align} \label{eq:momentumcurrent}
    p_{ab} = ( p - \phi_{cd} \varepsilon_{cd}) \delta_{ab} - \phi_{ab} + \frac{1}{2} (\phi_{ac} \varepsilon_{bc} + \phi_{bc} \varepsilon_{ac}).
\end{align}
For completeness, we rewrite Eqs.~\eqref{eq:n} to \eqref{eq:u} in real-space representation:
\begin{align} \label{eq:rn}
    &\partial_t n + \partial_a (n v_a) = 0,
    \\ \label{eq:rpi}
    &\partial_t \pi_a + \partial_b (\pi_a v_b + p_{ab} - \eta_{abdc} \partial_c v_d + N^{\pi_a}_b) = 0,
    \\ \label{eq:re}
    &\partial_t e + \partial_a (e v_a + p_{ab} v_b - \kappa_{ab} \partial_b T - \xi_{ab} \partial_c \phi_{bc} + N^{e}_a) = 0,
    \\  \label{eq:ru}
    &\partial_t u_a + v_b \partial_b u_a = v_a + \frac{\xi_{ba}}{T} \partial_b T + \zeta_{ab} \partial_c \phi_{bc} + N^{u_a}
\end{align}
with the white Gaussian noises obeying
\begin{align}
    \langle N^{e}_a (r,t) N^{e}_b (r^{\prime},t^{\prime}) \rangle &= 2 T_{\eq}^2 \kappa_{ab} \delta_{\Lambda} (r-r^{\prime}) \delta (t-t^{\prime}),
    \\
    \langle N^{e}_a (r,t) N^{\pi_a}_b(r^{\prime},t^{\prime}) \rangle &= 0,
    \\
    \langle N^{e}_a (r,t) N^{u_b} (r^{\prime}, t^{\prime}) \rangle &= 2 T_{\eq} \xi_{ab} \delta_{\Lambda}(r-r^{\prime}) \delta (t-t^{\prime}),
    \\
    \langle N^{\pi_a}_c (r,t) N^{\pi_b}_d (r^{\prime}, t^{\prime}) \rangle &= 2 T_{\eq} \eta_{acbd} \delta_{\Lambda} (r - r^{\prime}) \delta (t - t^{\prime}),
    \\
    \langle N^{\pi_a}_c (r,t) N^{u_b} (r^{\prime}, t^{\prime}) \rangle &= 0,
    \\
    \langle N^{u_a}(r,t) N^{u_b}(r^{\prime}, t^{\prime}) \rangle &= 2 T_{\eq} \zeta_{ab} \delta_{\Lambda} (r-r^{\prime}) \delta (t-t^{\prime}).
\end{align}
Here, $\delta_{\Lambda}$ is the mollified delta function defined by
\begin{align}
    \delta_{\Lambda} (r) \coloneqq \frac{1}{V} \sum_{k \in R_L^{\Lambda}} e^{ik \cdot r},
\end{align}
which recovers the standard delta function in the limit $V \to \infty$ and $\Lambda \to \infty$. The main results of this paper, Eqs.~\eqref{eq:rn} to \eqref{eq:ru}, are exactly the same as the hydrodynamic equation phenomenologically derived, except for the presence of the noises. See Appendix \ref{subsec:thermodynamics} for the phenomenological derivation of the hydrodynamic equation.

\section{Concluding remarks}
\label{sec:conclusion}

In summary, we have proposed a microscopic expression for the displacement field that correctly produces the nonlinear reversible decay rate and compactly derived the nonlinear fluctuating hydrodynamic equation for crystalline solids from the microscopic description of a many-particle system. We have provided thermodynamic expressions of the reversible current densities and decay rate that agree with those in the phenomenological hydrodynamic equation.

As a final remark, we note that the displacement field proposed in this paper describes only small fluctuations around a homogeneous crystal in equilibrium. This restriction originates from our method of defining the displacement field. When we define the displacement field through the relation (\ref{eq:nu}), we choose the homogeneous crystal as the reference state, i.e., $n_{\eq}(r) = \sum_{g \in G} n_{\eq,g} e^{ig \cdot r}$. Therefore, our result may not be applied to crystals in which the arrangement of particles is disturbed by dislocations. It would be challenging to determine the stochastic dynamics of crystals with topological defects from microscopic descriptions. Whether the hydrodynamic equations, Eqs.~\eqref{eq:rn} to \eqref{eq:ru}, describe the dynamics of a crystal under nonequilibrium conditions is also far from obvious because the crystals may exhibit large deformations even under the infinitesimally small external shear stress \cite{SaussetBiroliKurchan2010}. This is in contrast to fluctuating hydrodynamic equations for fluid systems \cite{LutskoDufty1985}. Developing the theory of nonequilibrium stochastic dynamics for crystals is left for future studies.

\section*{Acknowledgments}
I would like to thank Saswati Ganguly and Florian Miserez for kindly agreeing to send me a soft copy of Ref. \cite{Miserez2021}. This work was supported by JSPS KAKENHI Grant Number JP22J00337.

\appendix

\section{Preliminaries}

\subsection{Local thermodynamics for crystals}
\label{subsec:thermodynamics}

We review thermodynamic relations and phenomenological hydrodynamics for crystals. The contents of this section are found in many literatures \cite{MartinParodiPershan1972,FlemingCohen1976,Miserez2021}.

By introducing the internal energy density $e^* \coloneqq e - mn|v|^2/2$ and the chemical potential $\mu^* \coloneqq \mu + m|v|^2/2$ in the rest frame, we can rewrite the thermodynamic relation (\ref{eq:thr}) as
\begin{align}
    ds = \beta de^* - \beta \mu^* dn - \beta \phi_{ab} d \varepsilon_{ab}.
\end{align}
We now suppose that the conserved density fields and the displacement field obey the equations
\begin{align} \label{eq:cont}
    \partial_t n + \partial_a j_a^n = 0, \ \partial_t \pi_a + \partial_b j^{\pi}_{ab} = 0, \ \partial_t e + \partial_a j^e_a = 0,
\end{align}
and
\begin{align} \label{eq:dec}
    \partial_t u_a = - j_a^u,
\end{align}
where the specific forms of the conserved current densities $j^n_a$, $j^{\pi}_{ab}$, and $j^e_a$ and the decay rate $j^u_a$ will be determined later. From Eq.~\eqref{eq:thr}, the derivatives of the entropy functional (\ref{eq:calS0}) with respect to slow variables are
\begin{align} \label{eq:tf1}
    \frac{\delta \calS_0}{\delta e} = \beta, \ \frac{\delta \calS_0}{\delta \pi_a} = - \beta v_a, \ \frac{\delta \calS_0}{\delta n} = - \beta \mu
\end{align}
and
\begin{align} \label{eq:tf2}
    \frac{\delta \calS_0}{\delta u_a} = \partial_b (\beta \phi_{ab}) - \partial_b (\beta \phi_{bc} \partial_c u_a).
\end{align}
Using Eq.~\eqref{eq:thr}, we find that the entropy production rate for the solution of Eqs.~\eqref{eq:cont} and \eqref{eq:dec} is
\begin{align} \label{eq:ep}
    \frac{d}{dt} \calS_0 &= \int_{B_L} dx \left[ \partial_a \left( \frac{\delta \calS_0}{\delta e} \right) (j^e_a - ev_a - p_{ab} v_b) \right.
    \notag \\
    &\qquad + \partial_b \left( \frac{\delta \calS_0}{\delta \pi_a} \right) (j^{\pi}_{ab} - \pi_a v_b - p_{ab})
    \notag \\
    &\qquad \left. - \left( \frac{\delta \calS_0}{\delta u_a} \right) (j^u_a + v_a - v_b \partial_b u_a) + O((\delta \alpha)^4)  \right].
\end{align}
We decompose the current densities and decay rate into reversible and irreversible parts: $j^e_a = j^{e,\rev}_a + j^{e,\irr}_a$, $j^{\pi}_{ab} = j^{\pi,\rev}_{ab} + j^{\pi,\irr}_{ab}$, and $j^u_a = j^{u,\rev}_a + j^{u,\irr}_a$. If we require the three terms in Eq.~\eqref{eq:ep} to be zero for the reversible parts of the current densities and decay rate, then Eq.~\eqref{eq:ep} implies that
\begin{align}
    j^{e,\rev}_a &= ev_a + p_{ab} v_b,
    \\
    j^{\pi,\rev}_{ab} &= \pi_a v_b + p_{ab},
    \\ \label{eq:juph}
    j^{u,\rev}_a &= - v_a + v_b \partial_b u_a,
\end{align}
up to $O((\delta \alpha)^2)$. The positivity of the entropy production rate motivates us to write the irreversible parts as
\begin{align}
    j^{e,\irr}_a &= L^{ee}_{ab} \partial_b \left( \frac{\delta \calS_0}{\delta e} \right) - L^{e u_b}_{a*} \left( \frac{\delta \calS_0}{\delta u_b} \right),
    \\
    j^{\pi, \irr}_{ab} &= L^{\pi_a \pi_c}_{bd} \partial_c \left( \frac{\delta \calS_0}{\delta \pi_d} \right),
    \\
    j^{u, \irr}_a &= L^{u_a e}_{*b} \partial_b \left( \frac{\delta \calS_0}{\delta e} \right) - L^{u_a u_b}_{**} \left( \frac{\delta \calS_0}{\delta u_b} \right),
\end{align}
where $L^{ee}_{ab} = L^{ee}_{ba}$, $L^{eu_b}_{a*} = L^{u_be}_{*a}$, $L^{\pi_a \pi_c}_{bd} = L^{\pi_c \pi_a}_{db}$, and $L^{u_a u_b}_{**} = L^{u_b u_b}_{**}$ are the phenomenological Onsager coefficients. With the conventional choice of the transport coefficients in the main text, they are expressed as
\begin{align}
    j^{e,\irr}_a &= - \kappa_{ab} \partial_b T - \xi_{ab} \partial_c \phi_{bc},
    \\
    j^{\pi,\irr}_{ab} &= - \eta_{abcd} \partial_c v_d,
    \\
    j^{u,\irr}_a &= - \frac{\xi_{ba}}{T} \partial_b T - \zeta_{ab} \partial_c \phi_{bc},
\end{align}
up to $O(\delta \alpha)$ \cite{MartinParodiPershan1972,FlemingCohen1976}. Thus, the complete set of hydrodynamic equations for crystals is determined from a phenomenological argument.

\subsection{Microscopic expressions for locally conserved currents}
\label{subsec:microcurrent}

For readers' convenience, we summarize the microscopic expressions for locally conserved currents:
\begin{align}
    \hat{j}^n_{a,k} &= \frac{\hat{\pi}_{a,k}}{m},
    \\
    \hat{j}^{\pi}_{ab,k} &= \frac{1}{V} \sum_{i=1}^{N} \frac{p_{ia} p_{ib}}{m} e^{- ik \cdot r_i} 
    \notag \\
    &+ \frac{1}{2V} \sum_{i \neq j} F_a(r_i - r_j) (r_{ib} - r_{jb}) \frac{e^{- ik \cdot r_i} - e^{- ik \cdot r_j}}{-ik \cdot (r_i - r_j)},
    \\
    \hat{j}^e_{a,k} &= \frac{1}{V} \sum_{i=1}^{N} \frac{p_{ib}}{m} h_i e^{-ik \cdot r_i} 
    \notag \\
    &+ \frac{1}{2V} \sum_{i \neq j} \frac{p_{ia} + p_{ja}}{2m} F_a(r_i-r_j) (r_{ib} - r_{jb}) 
    \notag \\
    & \qquad \times \frac{e^{- ik \cdot r_i} - e^{- ik \cdot r_j}}{-ik \cdot (r_i - r_j)},
\end{align}
where $F_a (r) \coloneqq - \partial_a \Phi (r)$.

\section{Identification of displacement field}
\label{sec:disp}

\subsection{Motivation}
\label{subsec:motivation}

Let us consider a crystalline state with the equilibrium density profile $n_{\eq}$. For a small displacement field $u$, the number density field $n$ in the deformed crystal is
\begin{align} \label{eq:du}
    n(r) = n_{\eq}(r - u(r)) \det [ \delta_{ab} - \partial_a u_b].
\end{align}
The determinant on the right-hand side represents the change in the volume element due to the deformation. We require the microscopic displacement field $\hat{u}$ to satisfy Eq.~\eqref{eq:du} if we replace $n$ with the microscopic number density field $\hat{n}$. By expanding the relation up to $O(u^2)$, we obtain
\begin{align}
    \delta \hat{n} &= - \partial_a (n_{\eq} \hat{u}_a) + \frac{1}{2} (n_{\eq} \partial_a \hat{u}_a \partial_b \hat{u}_b - n_{\eq} \partial_a \hat{u}_b \partial_b \hat{u}_a 
    \notag \\
    &\quad + 2 \hat{u}_a \partial_a n_{\eq} \partial_b \hat{u}_b + \hat{u}_a \hat{u}_b \partial_a \partial_b n_{\eq}),
\end{align}
where $\delta \hat{n} \coloneqq \hat{n} - n_{\eq}$ is the deviation of the number density field from the equilibrium profile. Because we are concerned with the macroscopic displacement field, $\hat{u}$ is assumed to have only the Fourier components in the first Brillouin zone. Then, the above relation in momentum space becomes
\begin{align}
    \delta \hat{n}_{k+g} &= - i(k_a + g_a) n_{\eq,g} \hat{u}_{a,k} 
    \notag \\
    &\quad + \frac{1}{2} \sum_{q \in \BZ} n_{\eq,g} [ iq_a \hat{u}_{a,q} i(k_b-q_b) \hat{u}_{b,k-q}
    \notag \\
    &\qquad - iq_a \hat{u}_{b,q} i(k_b-q_b) \hat{u}_{a,k-q}
    \notag \\
    &\qquad + 2ig_a \hat{u}_{a,q} i(k_b-q_b) \hat{u}_{b,k-q}
    \notag \\
    &\qquad + ig_a ig_b \hat{u}_{a,q} \hat{u}_{b,k-q} ]
\end{align}
for $k \in \BZ$, where we have used $n_{\eq,k} = 0$ for $k \not\in G$. Multiplying both sides by $ig_c n_{\eq,g}^*$ and summing over $g \in G$, we obtain
\begin{align} \label{eq:ub}
    \hat{u}_{a,k} = \hat{u}^0_{a,k} + \sum_{q \in \BZ} \hat{u}_{a,q} i(k_b-q_b) \hat{u}_{b,k-q}.
\end{align}
Here, we have used $\sum_{g \in G} g_a |n_{\eq,g}|^2 = 0$. The iterative approximation of Eq.~\eqref{eq:ub} leads to Eq.~\eqref{eq:defdis}.

\subsection{Microscopic expression for decay rate}
\label{subsec:decay}

The decay rate $\hat{j}^{u^0}_{a,k}$ for the linear displacement field $\hat{u}_{a,k}^0$ is defined through the time evolution equation
\begin{align}
    \partial_t \hat{u}_{a,k}^0 = - \hat{j}^{u^0}_{a,k}(\Gamma_t).
\end{align}
From this, we obtain
\begin{align} \label{eq:ju0}
    \hat{j}^{u^0}_{a,k} = - (\calN^{-1})_{ab} \sum_{g \in G} g_b (g_c + k_c) n_{\eq,g}^* \frac{\hat{\pi}_{c, g+k}}{m}.
\end{align}
Therefore, the decay rate $\hat{j}^u_{a,k}$ for the displacement field $\hat{u}_{a,k}$ defined in Eq.~\eqref{eq:defdis} is
\begin{align}
    \hat{j}_{a,k}^u &= \hat{j}_{a,k}^{u^0} + \sum_{q \in R_L^{\Lambda}} (\hat{j}_{a,k}^{u^0} \cdot i(k_b-q_b) \hat{u}_{b,k-q}^0 
    \notag \\
    &\qquad + \hat{u}_{a,q}^0 \cdot i (k_b - q_b) \hat{j}_{b,k-q}^{u^0})
    \notag \\
    &\simeq \hat{j}_{a,k}^{u^0} + \sum_{q \in R_L^{\Lambda}} (\hat{j}_{a,k}^{u^0} \cdot i(k_b-q_b) \hat{u}_{b,k-q} 
    \notag \\ \label{eq:mexju}
    &\qquad + \hat{u}_{a,q} \cdot i (k_b - q_b) \hat{j}_{b,k-q}^{u^0}).
\end{align}
For simplicity, we have restricted the region of the wavenumbers to $R_L^{\Lambda}$. The local Galilean transformation for $\hat{j}^{u^0}_{a,k}$ is obtained as follows. Using Eq.~\eqref{eq:ju0} and \eqref{eq:ecm}, we find that
\begin{align}
    &\hat{j}^{u^0}_{a,k}
    \notag \\
    &= - (\calN^{-1})_{ab} \sum_{g \in G} g_b (g_c + k_c) n_{\mathrm{eq},g}^* \frac{\hat{\pi}_{c,g+k}^v + m (\hat{n}^v v_c)_{k+g}}{m}
    \notag \\
    &= \hat{j}^{u^0,v}_{a,k} + (\calN^{-1})_{ab} \sum_{g \in G} ig_b n_{\mathrm{eq},g}^* \cdot i (k_c + g_c)
    \notag \\
    &\qquad \times \sum_{q \in R_L^{\Lambda}} (n_{\mathrm{eq}, k+g-q} + \delta \hat{n}_{k-q+g}) v_{c,q}
    \notag \\
    &= \hat{j}^{u^0,v}_{a,k} - v_{a,k}
    \notag \\
    &+ \sum_{q \in R_L^{\Lambda}} (\calN^{-1})_{ab} \sum_{g \in G} ig_b n_{\mathrm{eq},g}^*  \delta \hat{n}_{k-q+g} \cdot i q_c v_{c,q}
    \notag \\
    &+ \sum_{q \in R_L^{\Lambda}} (\calN^{-1})_{ab} \sum_{g \in G} ig_b \cdot i (k_c - q_c + g_c) n_{\mathrm{eq},g}^*  \delta \hat{n}_{k-q+g} v_{c,q}
    \notag \\
    &= \hat{j}^{u^0,v}_{a,k} - v_{a,k} + \sum_{q \in R_L^{\Lambda}} \hat{u}^0_{a,k-q} \cdot i q_c v_{c,q}
    \notag \\
    &+ \sum_{q \in R_L^{\Lambda}} (\calN^{-1})_{ab} \sum_{g \in G} ig_b \cdot i (k_c - q_c + g_c) n_{\mathrm{eq},g}^*  \delta \hat{n}_{k-q+g} v_{c,q}
\end{align}
Here, we have used $n_{\mathrm{eq},k+g-q} = n_{\mathrm{eq},g} \delta_{k,q}$ for $k,q \in R_L^{\Lambda}$ and $\sum_{g \in G} g_b |n_{\mathrm{eq}}|^2 = 0$ in the third equality. We now recall that
\begin{align}
    \delta \hat{n}_{k+q} = - i (k_a + g_a) n_{\mathrm{eq},g} \hat{u}^0_{a,k}.
\end{align}
By multiplying both sides by $i(k_c+g_c) \cdot ig_b n_{\mathrm{eq},g}^*$ and summing over $g \in G$, we find that
\begin{align}
    &\sum_{g \in G} ig_b \cdot i(k_c + g_c) n_{\mathrm{eq},g}^* \delta \hat{n}_{k+g} 
    \notag \\
    &= i \left( \sum_{g \in G} g_b (k_c+g_c)(k_a+g_a) |n_{\mathrm{eq},g}|^2 \right) \hat{u}^0_{a,k}
    \notag \\
    &= i (k_a \calN_{cb} + k_c \calN_{ab}) \hat{u}^0_{a,k},
\end{align}
which implies that
\begin{align}
    &(\calN^{-1})_{ab} \sum_{g \in G} ig_b \cdot i(k_c - q_c+g_c) n_{\mathrm{eq},g}^* \delta \hat{n}_{k-q + g}
    \notag \\
    &= \delta_{ac} i(k_b-q_b) \hat{u}^0_{b,k-q} + i(k_c-q_c) \hat{u}^0_{a,k-q}.
\end{align}
Therefore, we get
\begin{align} \label{eq:lgju0}
    \hat{j}^{u^0}_{a,k} &= \hat{j}^{u^0,v}_{a,k} - v_{a,k} + \sum_{q \in R_L^{\Lambda}} \hat{u}^0_{a,k-q} \cdot iq_b v_{b,q}
    \notag \\
    &+ \sum_{q \in R_L^{\Lambda}} i(k_b-q_b) \hat{u}^0_{b,k-q} \cdot v_{a,q}
    \notag \\
    &+ \sum_{q \in R_L^{\Lambda}} i(k_c-q_c) \hat{u}^0_{a,k-q} v_{c,q}
\end{align}
in momentum space and
\begin{align}
    \hat{j}^{u^0}_{a} \simeq \hat{j}^{u^0,v}_a - v_a + (\nabla \cdot v) \hat{u}_a + (\nabla \cdot \hat{u}) v_a + (v \cdot \nabla) \hat{u}_a
\end{align}
in real space. By substituting Eq.~\eqref{eq:lgju0} into Eq.~\eqref{eq:mexju}, we obtain Eq.~\eqref{eq:jucm}

\section{Review of projection operator method}
\label{sec:reviewprojection}

\subsection{Derivation of Fokker--Planck equation}
\label{subsec:fp}

The probability density $\hat{\rho}_t$ on the phase space can be decomposed into a component that depends on the microscopic configurations through the slow variables and the fast remaining components, $\hat{\rho}_t = \calP \hat{\rho}_t + \calQ \hat{\rho}_t$. We suppose that the initial density function satisfies $\calQ \hat{\rho}_0 = 0$. From the Liouville equation (\ref{eq:liouville}) with the decomposition via the projection, we can express the time evolution equation for $p_t$ as
\begin{align} \label{eq:master}
 &\partial_t p_t (\alpha) = - \sum_l \frac{\partial}{\partial \alpha_l} (V_l (\alpha) p_t (\alpha))
 \notag \\
 &+ \sum_{l,m} \frac{\partial}{\partial \alpha_l} \int_0^t ds \int d\alpha^{\prime} \sqrt{\Omega (\alpha)} L_{lm} (\alpha, \alpha^{\prime} ; t-s) \sqrt{\Omega (\alpha^{\prime})} 
 \notag \\
 &\qquad \times \frac{\partial}{\partial \alpha_m^{\prime}} \frac{p_s (\alpha^{\prime})}{\Omega (\alpha^{\prime})}.
\end{align}
 The diffusion kernel is defined as 
\begin{align}
    &L_{lm}(\alpha, \alpha^{\prime} ; t) 
    \notag \\
    &= \sqrt{\frac{\Omega (\alpha)}{\Omega (\alpha^{\prime})}} \langle (e^{\calQ \calL t} [ \delta (\alpha - \hat{\alpha}) \calQ \calL \hat{\alpha}_l ]) ( \calQ \calL \hat{\alpha}_m ) \rangle^{\LM}_{\alpha^{\prime}}.
\end{align}
Because the projection $\calQ$ eliminates the fast degrees of freedom in observables, we expect that $L_{lm}(\alpha, \alpha^{\prime} ; t)$ decays quickly with a microscopic time $\tau_{\mathrm{micro}}$. During the short time $\tau_{\mathrm{micro}}$, the slow variables $\hat{\alpha}$ and $p_t(\alpha)$ hardly change. Hence, we can approximate via $L_{lm}(\alpha, \alpha^{\prime} ;t) \simeq L_{lm}(\alpha ; t) \delta (\alpha - \alpha^{\prime})$ with $L_{lm}(\alpha ; t) \coloneqq \langle (e^{\calQ \calL t} \calQ \calL \hat{\alpha}_l)( \calQ \calL \hat{\alpha}_m) \rangle^{\LM}_{\alpha}$, and $p_s(\alpha^{\prime}) \simeq p_t (\alpha)$ in Eq.~\eqref{eq:master}. Moreover, we can safely extend the upper limit of the integral in Eq.~\eqref{eq:master} to the infinity for the same reason. Consequently, we obtain the Markovian Fokker--Planck equation (\ref{eq:fp}).

\subsection{Detailed balance condition}
\label{subsec:db}

The time-reversal symmetry of the Hamiltonian dynamics implies that the diffusion kernel satisfies the reciprocal relations $L_{lm}(\alpha, \alpha^{\prime} ; t) = \epsilon_l \epsilon_m L_{ml}(\epsilon \alpha^{\prime}, \epsilon \alpha ; t)$. In the Markovian limit, these are simplified to
\begin{align} \label{eq:reciprocal}
    L_{lm}(\alpha) = \epsilon_l \epsilon_m L_{ml}(\epsilon \alpha)
\end{align}
The drift vector of the Fokker--Planck equation (\ref{eq:fp}) is
\begin{align}
    F_l  = V_l + \sum_m L_{lm} \frac{\partial \calS}{\partial \alpha_m} + \sum_m \frac{\partial L_{lm}}{\partial \alpha_m}.
\end{align}
Its reversible and irreversible parts are, respectively,
\begin{align} 
    F^{\rev}_l (\alpha) &= \frac{1}{2} (F_l(\alpha) - \epsilon_l F_l (\epsilon \alpha))
    \\ \label{eq:rev}
    &= V_l (\alpha) + \sum_m \frac{1}{\Omega (\alpha)} \frac{\partial}{\partial \alpha_m} (L_{lm}^{(\mathrm{a})}(\alpha) \Omega (\alpha)) 
\end{align}
and
\begin{align} 
    F^{\irr}_l(\alpha) &= \frac{1}{2} (F_l (\alpha) + \epsilon_l F_l (\epsilon_l \alpha))
    \\ \label{eq:irr}
    &= \sum_m L_{lm}^{(\mathrm{s})}(\alpha) \frac{\partial \calS (\alpha)}{\partial \alpha_m} + \sum_m \frac{\partial L_{lm}^{(\mathrm{s})}(\alpha)}{\partial \alpha_m},
\end{align}
where $L_{lm}^{(\mathrm{a})}(\alpha) = (L_{lm}(\alpha) - L_{ml}(\alpha))/2$ is the anti-symmetric part of the Onsager matrix. We can prove that the reversible and irreversible drifts satisfy
\begin{align} \label{eq:irrdb}
    F_l^{\irr} \Omega - \sum_m \frac{\partial }{\partial \alpha_m} (L_{lm}^{(\mathrm{s})} \Omega) = 0
\end{align}
and
\begin{align} \label{eq:stationary}
    \sum_l \frac{\partial}{\partial \alpha_l} (F^{\rev}_l \Omega) = 0.
\end{align}
The above two properties and the reciprocal relation (\ref{eq:reciprocal}) are equivalent to the detailed balance condition \cite{GrahamHaken1971}. The second property (\ref{eq:stationary}) is simply the stationarity condition of the Fokker--Planck equation for $\Omega$. Eq.~\eqref{eq:irrdb} is obvious from Eq.~\eqref{eq:irr}. The proof of Eq.~\eqref{eq:stationary} is as follows:
\begin{align*}
    \sum_l \frac{\partial}{\partial \alpha_l} (V_l \Omega) &= \sum_l \frac{\partial}{\partial \alpha_l} \Tr [\{ \hat{\alpha_l}, \hat{H} \} \delta (\alpha - \hat{\alpha})]
    \notag \\
    &= - \sum_l \frac{\partial}{\partial \alpha_l} \Tr [ \{ \hat{\alpha}_l, \delta (\alpha - \hat{\alpha}) \} \hat{H} ]
    \notag \\
    &= \sum_{l,m} \frac{\partial}{\partial \alpha_l} \frac{\partial}{\partial \alpha_m} \Tr [ \{ \hat{\alpha}_l, \hat{\alpha}_m \} \delta (\alpha - \hat{\alpha}) \hat{H}  ] 
    \notag \\
    &= 0
\end{align*}
and
\begin{align}
    \sum_{l,m} \frac{\partial}{\partial \alpha_l} \frac{\partial}{\partial \alpha_m} (L_{lm}^{(\mathrm{a})} \Omega) = 0
\end{align}
from the anti-symmetry of $L_{lm}^{(\mathrm{a})}$.


\end{document}